\documentclass[sigconf]{acmart}
\usepackage{bm}
\usepackage{amsmath}
\usepackage{amssymb}
\usepackage{booktabs}
\usepackage{graphicx}
\usepackage{caption}
\usepackage{subcaption}
\AtBeginDocument{%
  \providecommand\BibTeX{{%
    \normalfont B\kern-0.5em{\scshape i\kern-0.25em b}\kern-0.8em\TeX}}}

\copyrightyear{2020}
\acmYear{2020}
\setcopyright{acmcopyright}\acmConference[SIGIR '20]{Proceedings of the 43rd International ACM SIGIR Conference on Research and Development in Information Retrieval}{July 25--30, 2020}{Virtual Event, China}
\acmBooktitle{Proceedings of the 43rd International ACM SIGIR Conference on Research and Development in Information Retrieval (SIGIR '20), July 25--30, 2020, Virtual Event, China}
\acmPrice{15.00}
\acmDOI{10.1145/3397271.3401044}
\acmISBN{978-1-4503-8016-4/20/07}

\begin{document}
\fancyhead{}
\title{Table Search Using a Deep Contextualized Language Model}

\author{Zhiyu Chen}
\email{zhc415@lehigh.edu}
\affiliation{%
  \institution{Lehigh University}
  \streetaddress{113 Research Drive (Building C)}
  \city{Bethlehem}
  \state{PA}
  \country{USA}
  \postcode{18015}
}

\author{Mohamed	Trabelsi}
\email{mot218@lehigh.edu}
\affiliation{%
  \institution{Lehigh University}
  \streetaddress{113 Research Drive (Building C)}
  \city{Bethlehem}
  \state{PA}
  \country{USA}
  \postcode{18015}
}

\author{Jeff Heflin}
\email{heflin@cse.lehigh.edu}
\affiliation{%
  \institution{Lehigh University}
  \streetaddress{113 Research Drive (Building C)}
  \city{Bethlehem}
  \state{PA}
  \country{USA}
  \postcode{18015}
}

\author{Yinan Xu}
\email{yinanxu@wezhuiyi.com}
\affiliation{%
  \institution{Zhuiyi Technology}
  \city{Shenzhen}
  \country{China}
  \postcode{18015}
}

\author{Brian D.\ Davison}
\email{davison@cse.lehigh.edu}
\affiliation{%
  \institution{Lehigh University}
  \streetaddress{113 Research Drive (Building C)}
  \city{Bethlehem}
  \state{PA}
  \country{USA}
  \postcode{18015}
}


\renewcommand{\shortauthors}{Zhiyu et al.}


\begin{abstract}
Pretrained contextualized language models such as BERT have achieved impressive results on various natural language processing benchmarks. Benefiting from multiple pretraining tasks and large scale training corpora, pretrained models can capture complex syntactic word relations. 
In this paper, we 
use the deep contextualized language model BERT for the task of ad hoc table retrieval. We investigate how to encode table content considering the table structure and input length limit of BERT. We also propose an approach that incorporates features from prior literature on table retrieval and jointly trains them with BERT. In experiments on public datasets, we show that our best approach can outperform the previous state-of-the-art method and BERT baselines with a large margin under different evaluation metrics. 
\end{abstract}

\ccsdesc[500]{Information systems~Content analysis and feature selection}
\ccsdesc[500]{Information systems~Retrieval models and ranking}
\ccsdesc[500]{Computing methodologies~Search methodologies}

\keywords{table search; neural networks; pretrained language model; information retrieval}


\maketitle
\section{Introduction}
As an efficient way to organize and display data, tables are broadly used in different applications: researchers use tables to present their experimental results; companies store information about customers and products in spreadsheets;  flight information display systems in the airports show flight schedules to passengers in tables. 
According to Cafarella et al.~\cite{cafarella2008webtables}, there are more than 14.1 billion tables on the Web.  Among those tables, many are very informative which means they include relations and attributes of real-world entities, and have been used for a variety of downstream tasks. 
For example, tables like Wikipedia infoboxes have been used to construct knowledge bases since they are of high quality and consistent structure \cite{auer2007dbpedia}. 
Data-to-text models take tables from specific domains as input and transform them into fluent natural language sentences such as sports news \cite{wiseman2017challenges}
and product descriptions \cite{chan-etal-2019-stick}.
With structure information and metadata, tables store factual knowledge and therefore are also used to build question answering (QA) systems \cite{sun2016table}
.  

\begin{figure}
	\centering
	\begin{subfigure}{0.385\textwidth} 
		\includegraphics[width=\textwidth]{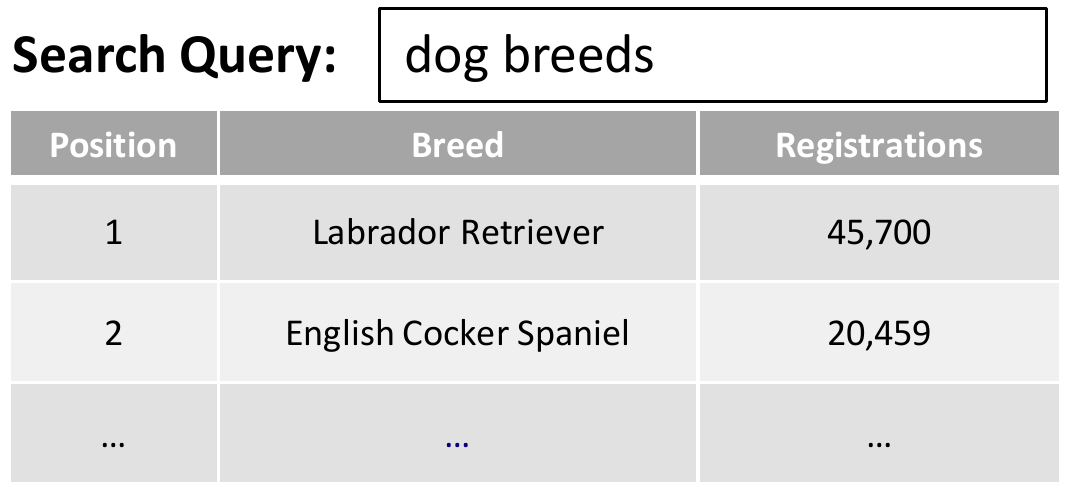}
		\caption{An example of a returned table in which one column is relevant to the query.} 
		\label{col_intent}
	\end{subfigure}
	\vspace{1em} 
	\begin{subfigure}{0.385\textwidth} 
		\includegraphics[width=\textwidth]{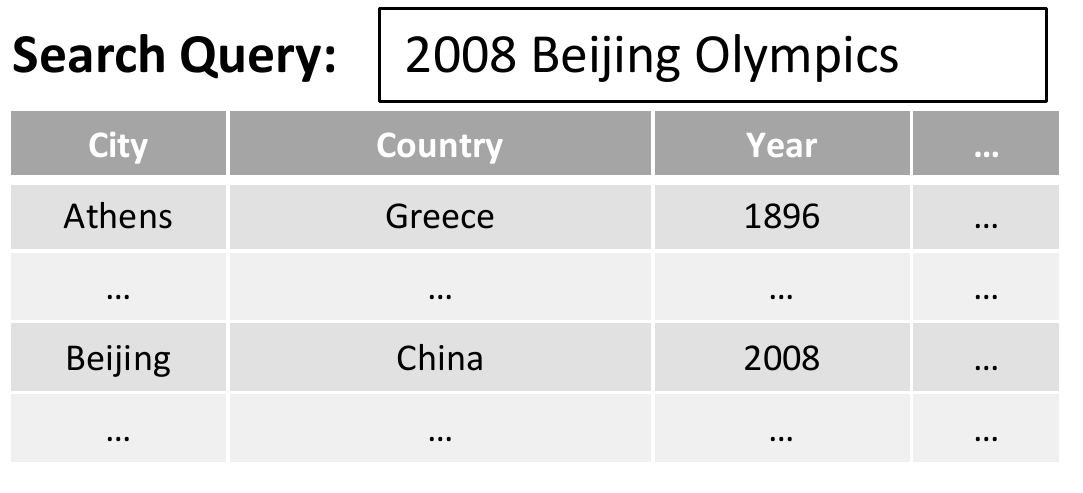}
		\caption{An example of a returned table in which one row is relevant to the query.}
    \label{row_intent}
	\end{subfigure}
	\begin{subfigure}{0.385\textwidth} 
		\includegraphics[width=\textwidth]{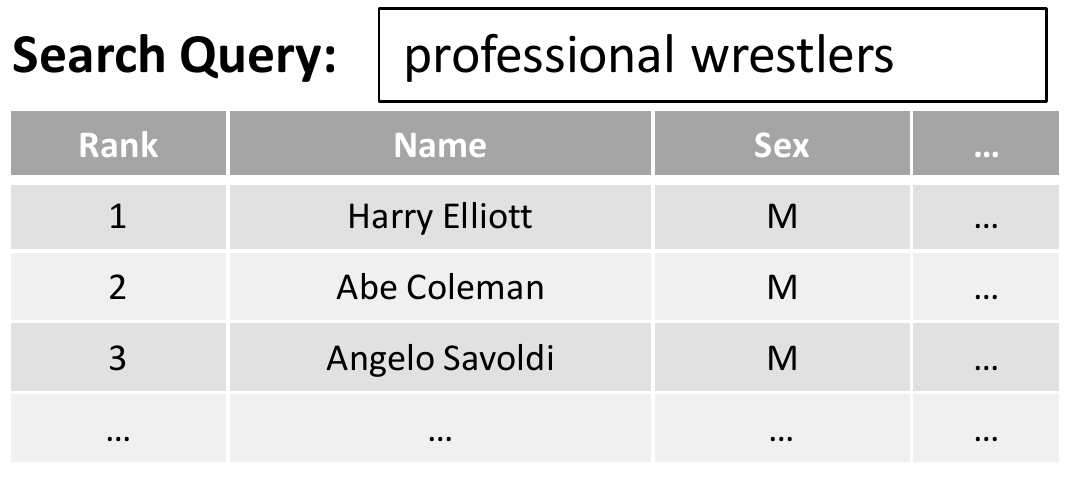}
		\caption{An example of a returned table in which all cells are relevant to the query.}
        \label{cell_intent}
	\end{subfigure}
	\caption{Three examples of returned tables reflecting different relevant unit types.} 
	\label{answer_types}
\end{figure}

The table retrieval task is related but different from the table QA task. Both of them aim to satisfy users' information need. QA models usually take a natural language question as input and aim to find one or more specific answers. However, queries for table retrieval systems may have ambiguous intent and usually consist of several keywords. 
The returned tables from a table retrieval system in Figure \ref{col_intent} and \ref{cell_intent} can be both positive samples for a table QA system. A user may want to know the list of all dog breeds in Figure \ref{col_intent} and the 2nd column of the table provides the relevant and accurate information. A user may ask the profiles of professional wrestlers in Figure \ref{cell_intent} and the returned table contains that information. For this example, all the cells provide informative content for the user. The 2nd column tells who are professional wrestlers and the other columns provide context information.
However, in Figure \ref{row_intent}, the query has more ambiguous intent. The user may ask the results of 2008 Beijing Olympic Games which means the returned table is a negative sample for a QA system. If a user does not have a clear question and just wants to explore what he/she could find, then the returned table is a positive sample for a table retrieval system. For this example, the row that includes ``Beijing'' is relevant and the remaining rows are less useful.
We note that the unit of relevant information in the table can be rows, columns or cells. Based on this observation, we propose different methods to select row items, column items and cell items from a table.

In this paper, we consider the task of ad hoc table retrieval where given a keyword query, a list of ranked tables are returned.
In previous studies of table retrieval, various features are used. Word level, phrase level and sentence level features are calculated by Sun et al.~\cite{sun2019content}. Zhang et al.~\cite{zhang2018ad} use 23 hand-crafted features and  16 embedding based features to train a random forest for pointwise table ranking.
Recently, the pre-trained language model BERT~\cite{devlin2018bert} and its variants like RoBERTa~\cite{liu2019roberta} have achieved impressive results on different natural language understanding tasks~\cite{wang2019glue}. The self-attention structure and pre-training tasks enable BERT to learn complex linguistic features from a large corpus. Researchers from IR communities have applied BERT to ranking tasks and achieved new state-of-the-art results on multiple benchmarks~\cite{yang2019end,yang2019simple,MacAvaney:2019:CCE:3331184.3331317,nogueira2019multi}. 
Here we apply BERT to the ad hoc table retrieval task.
In previous work, the input of BERT is either a single sequence or sequence pairs. 
The question of how to effectively encode a structured document into a BERT representation has not been previously explored.  
We construct input for BERT considering the structure of a table and then combine BERT features with other table features together to 
treat table retrieval as a regression task.

We summarize our contributions as the following:
\begin{itemize}
    \item We propose three content selectors to encode different table items into the fixed-width BERT representation.
    \item We experiment on two 
    public 
    datasets and demonstrate that our method 
    achieves the best results and generalizes to other domains.
    \item We analyze the experiment results and discuss why the max salience selector for row items performs the best among all other methods.
    \item We analyze the fine-tuned BERT attention maps and embeddings, and explain what information is captured by BERT.
\end{itemize}
\section{Related Work}

\subsection{Table Search}
Zhang et al.~\cite{zhang2018ad} propose a semantic table retrieval (STR) method for ad hoc table retrieval. They first map queries and tables into a set of word embeddings or graph embeddings. Four ways to calculate query-table similarity based on embeddings are then proposed. In the end, the resulting four semantic similarity features are combined with other features into a learning-to-rank framework.  Table2Vec~\cite{zhang2019table2vec}  obtains semantic features in a similar way but uses embeddings trained from different fields.  This method is built upon and does not outperform STR, so we only compare our methods with STR instead of Table2Vec.

Unsupervised methods for table search are also studied. 
Trabelsi et al.~\cite{mo2019bdt}  propose custom embeddings for column headers based on multiple contexts for table retrieval, and find representing numerical cell values to be useful.
Chen et al.~\cite{chen2019smlr} utilize matrix factorization to generate additional table headers and then show that those generated headers can improve the performance of unsupervised table search.

\subsection{Retrieval Models for Multifield Documents}

A table is often associated with important context information such as its caption and can be considered as a multifield document. Therefore, table search can be treated as a multifield document retrieval task and we introduce some related work in the area of multifield document ranking.

Considering the structure of a document when designing retrieval models can usually improve retrieval results. 
It has been shown that combining similarities and rankings of different sections can lead to better performance \cite{wilkinson1994effective}. 
Ogilvie et al.~\cite{ogilvie2003combining} present a mixture-based language model combining different document representations for known-item search in structured document collections. They find that document representations that perform poorly can be combined with other representations to improve the overall performance.
Robertson et al.~\cite{robertson2004simple} introduce BM25F which is an extension of BM25 that combines original term frequencies in the different fields in a weighted manner.
A field relevance model is proposed by Kim and Croft \cite{kim2012field}  to incorporate relevance feedback for field weights estimation.
There are also supervised methods for multifield document ranking. 
A Bayesian networks-based model for structured documents is proposed by Piwowarski and Gallinari \cite{piwowarski2003machine}.
Kim et al.~\cite{kim2009probabilistic} propose a probabilistic model for semi-structured document retrieval. They calculate the mapping probability of each query term and use it as a weight to combine the language models estimated from each field.
Svore et al.~\cite{svore2009machine} develop LambdaBM25, a machine learning approach to BM25-style retrieval that learns from the input attributes of BM25 and performs better than BM25F for multifield document ranking.
Zamani et al.~\cite{zamani2018neural} propose a  neural ranking model that learns an aggregated document representation from field-level representations and then uses a matching network to produce the final relevance score.


\subsection{BERT for Information Retrieval}

Given the advances of deep contextualized language models for natural language understanding tasks, researchers from IR community also begin to study BERT for IR problems.
Nogueira et al.~\cite{nogueira2019passage} describe an initial application of BERT for passage re-ranking task where the sentence-pair classification score is used.
Nogueira et al.~\cite{nogueira2019multi} then propose a multi-stage document ranking framework where BERT is used for pointwise and pairwise re-ranking.
Yang et al.~\cite{yang2019simple} show that treating social media text retrieval as a sentence pair classification task can achieve new state-of-the-art results. Then they apply BERT to a dataset with longer documents and rank a document with linear interpolation of the original document score and weighted top-n sentence scores. 
Similarly, Dai et al.~\cite{dai2019deeper} use passage-level evidence to fine-tune BERT and consider all passages from a relevant document as relevant.  They first predict the relevance score of each passage independently. The document relevance is  the score of the first passage, the best passage, or the sum of all passage scores. 
BERT has also been applied to FAQ retrieval task by Sakata et al~\cite{faqbert} where given a user query, a question is scored by the combination of question-question BM25 score and question-answer BERT score.
MacAvaney et al.~\cite{MacAvaney:2019:CCE:3331184.3331317} combine the BERT classification token with existing neural IR models. 
The experiments show that this joint approach can outperform a vanilla BERT ranker.

IR researchers also investigate the possible reasons why BERT can have such substantial improvements for IR problems.
Through carefully designed experiments, Padigela et al.~\cite{padigela2019investigating} show that BM25 is more biased towards high-frequency terms which hurt its performance while BERT has a better ability to discover the semantic meaning of novel terms in documents with respect to query terms. They also find that BERT has less performance improvement compared with BM25 for long queries.
Dai et al.~\cite{dai2019deeper} demonstrate that BERT can take  advantage of stopwords and punctuation in the queries which is in contrast to traditional IR models. 
Qiao et al.~\cite{qiao2019understanding} show that BERT can be categorized into interaction-based IR models because simply obtaining query and document  representations from BERT independently and then computing their cosine similarity results in performance close to random. They also find that BERT assigns extreme matching scores to query-document pairs and most pairs get either one or zero ranking scores.

Many researchers (e.g.,  \cite{nogueira2019multi,yang2019simple,dai2019deeper,mass2019study}) find that the length limit of BERT causes difficulties in training. Mass et al.~\cite{mass2019study} specifically study the effect of passage length and segmentation configurations on passage re-rank performance. They find that mid-sized (256 tokens) inputs achieve the best results for the selected datasets. Dai and Callan's method \cite{dai2019deeper} to deal with long documents as mentioned before may result in noisy positive samples because for a relevant document, not all sentences are relevant to a query. The splitting and then aggregating methods in these approaches can increase the training and inference cost several times. In this paper, we pre-select the segments from the input with low-cost methods and then use BERT for the downstream table retrieval task.

\section{Prerequisites}

\subsection{BERT}\label{bert}

BERT~\cite{devlin2018bert}, consisting of $L$ layers of Transformer blocks, is a deep contextual language model which has achieved impressive results on various natural language processing tasks. Given a sequence of input token embeddings $\bm{X}=\{\bm{x}_1, \bm{x}_2, ..., \bm{x}_n\}$, the Transformer block at layer $l$ outputs the contextualized embeddings (hidden states) of input tokens $\bm{H}^l=\{\bm{h}_1^l, \bm{h}_2^l, ..., \bm{h}_n^l\}$. The Transformer block is originally proposed by Vaswani et al.~\cite{vaswani2017attention} and each has the same structure: multi-head self-attention followed by a feed-forward network. 
\begin{equation}
    \begin{aligned}
        &Transformer_l(\bm{H}^{l-1})\\
        &=FFN(MH\_Attn(\bm{H}^{l-1}))\\
        &=FFN(\bm{W}[Attn_1(\bm{H}^{l-1}), ..., Attn_m(\bm{H}^{l-1})])
    \end{aligned}
\end{equation}
Multi-head self-attention aggregates the output from $m$ attention heads.

When using BERT for downstream tasks, special tokens ([SEP] and [CLS]) are added into the input. For single sequence classification/regression tasks,  [CLS] and [SEP] are added to the beginning and end of the input sequence. For sequence-pair classification/regression, the two input sequences are concatenated by [SEP] and then processed the same as single sequence tasks. The embedding of [CLS] from the last Transformer block is fed into a final classification/regression layer. 

\subsection{BERT Characteristics} 

\textbf{Limit on input length.} BERT cannot take input sequences longer than 512 tokens. In previous studies of BERT for long document tasks like text classification \cite{chang2019xbert},  the input tokens are truncated. Better ways to preprocess the inputs beyond length limitation  are worth studying since trivially throwing away part of the inputs could lose important information. Transformer-XL~\cite{dai2019transformer} solves the fixed-length issue with recursion and relative position encoding. However, this method requires further pre-training and is only evaluated on text generation tasks. Though we focus on table retrieval, our methods to alleviate the long sequence issue are off-the-shelf without any further training and can also be applied to other domains. 

\textbf{The secrets behind special tokens.}  Before BERT was proposed, neural models for NLP and IR tasks usually take the embeddings of all input tokens for training. While for BERT and its variants, fine-tuning on the target tasks only requires an additional softmax layer on the top of the [CLS] embedding from the last layer and the remaining embeddings are not used. The function of [SEP] is often disregarded, as when constructing the input of BERT, the role of [SEP] is just a symbol to mark the end or delimiter of a sequence. 
Recently, researchers begin to analyze why BERT is so effective for different tasks. 
Clark et al.~\cite{clark2019what} suggest that [SEP] might be used as a "no-op" sometimes and does not aggregate segment-level information. 
However, Ma et al.~\cite{ma2019universal} show that using the embedding of [SEP] instead of [CLS] can also achieve comparable results, which indicates that [SEP] also learns contextualized information of the sequence.  
In our experiments, we  study the relationship  between special tokens and other input tokens in order to explore what BERT embeddings learn after fine-tuning on the target task.

\section{Method}

Here we define the task and then describe our method in detail.

\subsection{Task Definition}

In ad hoc table retrieval, given a query $q\in Q$ usually consisting of several keywords $q=\{k_1,k_2,...,k_l\}$, our goal is to rank a set of tables $T=\{t_1,t_2,...,t_n\}$ in descending order of their relevance scores with respect to $q$. A table is a set of cells arranged in rows and columns like a matrix. Each cell could a be single word, a real number, a phrase or even sentences. The first row of a table is the header row and consists of header cells. In practice, tables from the Web could have more complex structures~\cite{crestan2011web}. In this paper, we only consider tables that have the simplest structure  since they are the most commonly used. Each table could have context fields $\{p_1,...,p_k\}$ depending on the source of the table. For example, a table from Wikipedia can have a caption, its section title and page title. 

\begin{figure}[ht!]
\centering
\includegraphics[width=0.45\textwidth]{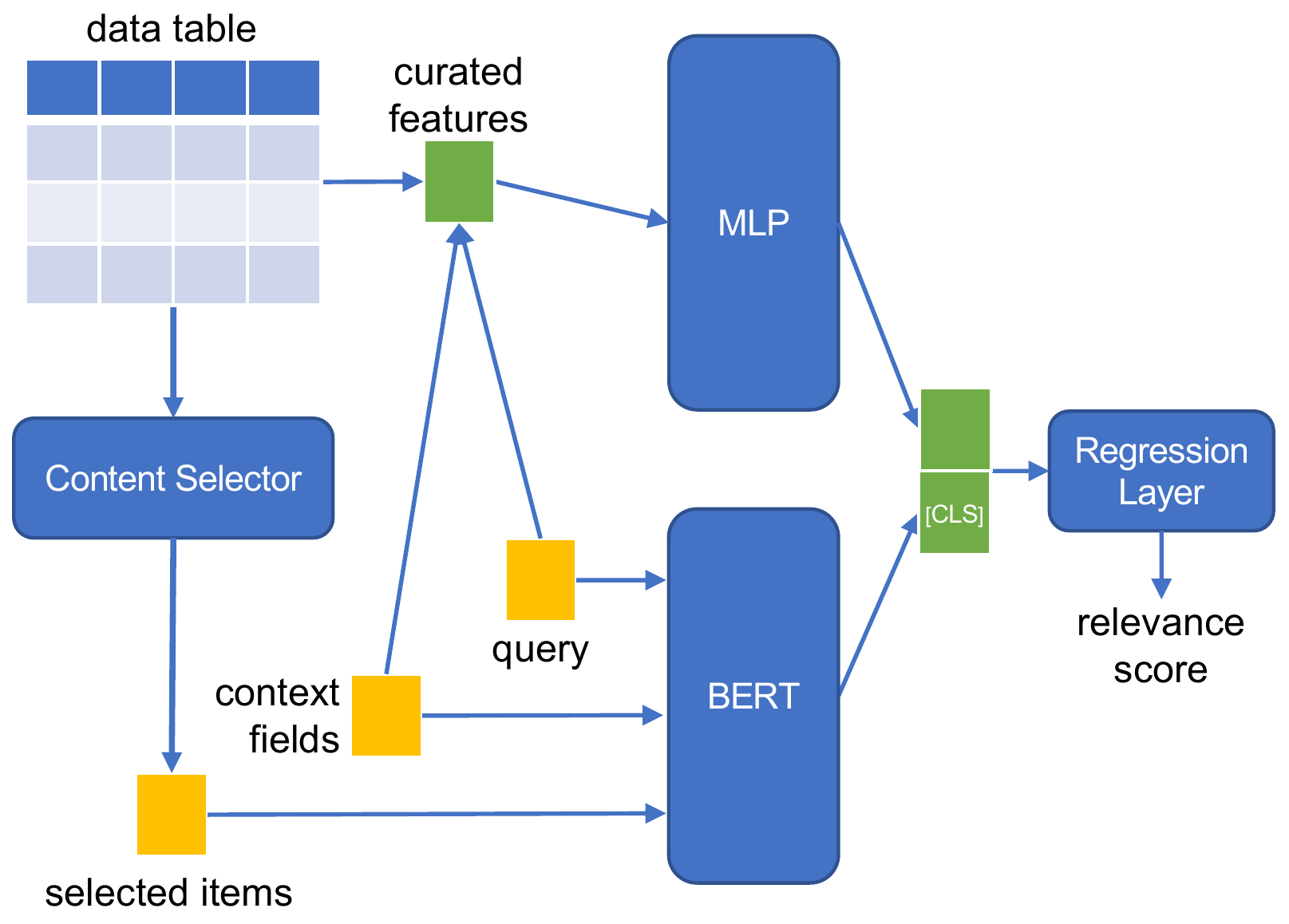}
\caption{Overview of the proposed model. Blue blocks are model components. Orange blocks are raw text of the input. Green blocks are either manually curated features or outputs from models (BERT and MLP).}
\label{method}
\end{figure}

\subsection{BERT for Table Retrieval}

We show the overview of our framework which includes four components in Figure~\ref{method}. 
The content selector extracts informative items (rows, columns or cells) from a table.
BERT is used to extract features $f_{bert}$ from the query, corresponding table context fields, and selected items. 
A neural network is used to transform additional features $v_a$(if provided) to $f_{a}$.
Then $f_{bert}$ and $f_{a}$ are concatenated into a single feature vector. This vector is fed into a regression layer to predict the  relevance score. In the rest of this section, we describe the model components in detail.

\subsubsection{Content Selector}\label{select}

As previously mentioned, BERT can only take input sequences that are no longer than 512 tokens. But for many tasks including table retrieval, the lengths of inputs can easily exceed that limit. To deal with the limit for various downstream tasks, inputs are typically truncated into valid lengths or multiple instances for a single document are created~\cite{47761,yang2019end}. In open-domain question answering and machine reading comprehension,  a single instance usually involves long documents or multiple documents, and the proposed methods usually have a select-then-extract two-stage schema \cite{wang-etal-2018-joint,hu2019retrieve,lin2018denoising,lee-etal-2019-latent,wang-etal-2019-multi}. 
Inspired by those works, we propose a \textbf{select-then-rank} framework for ad hoc table retrieval. First, we select a set of potentially informative items from a table. Then we pack the context fields of a table and its selected items as the final table representation. Finally we extract BERT features $f_{bert}$  for all the tables based on the new constructed representation.

In the ad hoc table retrieval task, we notice that there are three types of relevant tables in terms of the unit of the relevant information: 
\begin{itemize}
    \item One or more columns are relevant to the query. For example, only the second column is relevant to the query in Figure \ref{col_intent}; 
    \item One or more rows are relevant to the query. For example, only the row that includes ``Beijing''  is relevant to the query in Figure \ref{row_intent}.
    \item The relevant information is spread over the whole table. For example, in Figure \ref{cell_intent}, the table includes a list of records about the entities asked by the query.
\end{itemize}

Therefore, we slice a table $t$ into a list of items $\{c_1,...,c_m\}$, i.e.,  a list of rows, columns or cells and  select the top-ranked items for the final BERT input representation. Here we propose three methods to measure the salience score of a table item $c$:

\begin{itemize}
    \item Mean Salience: it assumes that the relevance signal can be captured by the similarity of query representation and item representation. We use the average word embeddings to represent queries and items respectively.  
    $$SAL_{mean}(c) = cosine(\frac{\sum_{w \in c}v_w}{l_c},\frac{\sum_{k \in q}v_k}{l_q})$$ 
    \item Sum Salience: it assumes  relevance signals between every pair of query and item terms are useful for content selection.  
    $$SAL_{sum}(c) =\sum_{k \in q}\sum_{w \in c} cosine(v_k,v_w) $$ 
    \item Max Salience: it assumes that only the most salient signal between any pair of query and item terms is useful for content selection. 
    $$SAL_{max}(c) =\max_{k \in q, w \in c} cosine(v_k,v_w) $$ 
\end{itemize}

\noindent
Instead of trivially truncating table information, we rank the items of a table and keep items with higher salience scores in the front. 

\subsubsection{Encoding Table for BERT}\label{encodingtable}

Given a query $q \in Q$, a table $t \in T$, the context fields $\{p_1,...,p_k\}$ and selected items of $t$ $\{c_1,...,c_m\}$, we construct the final input sequence for BERT as 
$$S=[[CLS],q,[SEP],p_1,[SEP],...,p_k,[SEP],c_1,[SEP],...,c_m,[SEP]]$$ 
Like Hu et al.~\cite{devlin2018bert}, we use WordPiece tokenization for input sequences and the input representation of each token is constructed by summing its  token embedding, segment embedding and position embedding.  All the queries share the same segment embedding and context fields, selected items share another segment embedding. As illustrated in Section \ref{bert}, we use the embedding of $[CLS]$ from the last layer as BERT features $f_{bert}$.

\subsubsection{Feature Fusion and Prediction}

Feature-based methods have shown impressive performance and achieved previous state-of-the-art results on ad hoc table retrieval \cite{zhang2018ad}. When additional feature $v_a \in \mathbb{R}^d$ for a query-table pair is available, we combine them with BERT features $f_{bert}$ by:
\begin{equation}
    f_a =v_a\bm{W}_1 + b_1
\end{equation}
where $\bm{W}_1 \in \mathbb{R}^{d \times d} $.
Then $f_a$ and $f_{bert}$ are concatenated into single vector and fed to the final regression layer.
\begin{equation}\label{fusion}
    f = [f_a;f_{bert}]
\end{equation}
\begin{equation}
    score = Regression(f)
\end{equation}
When only BERT features are available, $f$ equals $f_{bert}$. 
A simple linear transformation is used as regression layer which means $Regression(f)=f\bm{W}_2 + b_2$ where $\bm{W}_2 \in \mathbb{R}^{(d+h) \times 1}$ and $h$ is the size of BERT hidden states. 

\subsubsection{Training}

We use the pre-trained BERT-large-cased model which consists of 24 layers of Transformer blocks, 16 self-attention heads per layer and has a hidden size of 1024. Considering processing speed, the size of GPU memory, and the fact that BERT is good for short text tasks, the maximum input length is set to 128. Since the selected items are at the end of the input (as described in Section \ref{encodingtable}) and ranked by their salience scores with respect to the query, we assume the truncated part will have the least negative impact with a given length constraint. Considering the dataset statistics in Table~\ref{wikitable}, we limit the caption to 20 tokens, section title and page title to 10 tokens each, and table headers to 20 tokens. Since queries are short, we keep all the query tokens. As a result, we leave about half of the space for table content.
We fine-tune the framework by minimizing the Mean Square Error (MSE) between model predictions and gold standard relevance scores.\footnote{We also tried binary classification to predict relevance probabilities as in Sakata et al.~\cite{10.1145/3331184.3331326} and found that regression is much better in our scenario.}
We train the model with 5 epochs and batch size of 16. The Adam optimizer with learning rate of 1e-5 is used. We also use a linear learning rate decay schedule with warm-up of 0.1.
Our implementation is based on code from an open source repository.\footnote{\url{https://github.com/huggingface/transformers}} 
\section{Experiments}

In this section, we aim to answer the following research questions:
\begin{itemize}
    \item[RQ1:] What is the performance gain of BERT with content selection methods, with respect to state-of-the-art performance?
    \item[RQ2:] Could BERT with content selection methods outperform state-of-the-art performance without additional features?
    \item[RQ3:] Which content selection method/item type is the most effective?
\end{itemize}

\subsection{Dataset Description}

We use the WikiTables dataset created by Zhang and Balog \cite{zhang2018ad}  where the previous state-of-the-art method is proposed. 
The table corpus is originally extracted from Wikipedia~\cite{bhagavatula2015tabel}. The context fields include page title and section title. From Figure \ref{row_intent} and Figure \ref{col_intent} we can see that the first row of a table usually contains some high-level concepts and provides informative context. Therefore we also consider the table header as a context field. When slicing the tables, we still have table headers included.
The queries are sampled from the collections in \cite{cafarella2009data,venetis2011recovering}. 
In total, they annotated 3120 query-table pairs. 
The statistics of the corpus are shown in Table \ref{wikitable}.
We also use the curated features proposed by Zhang and Balog \cite{zhang2018ad} for feature fusion.

\begin{table}[]
\centering
\caption{The length statistics of data provided by Zhang and Balog \cite{zhang2018ad}. The length is calculated after WordPiece tokenization.}
\label{wikitable}
\begin{tabular}{rrrcc}
\hline
Field & Mean  & Max  & > 512 & > 128 \\ \hline
query & 3.5 & 8 & - & -\\
caption & 4.3 & 76 & - & -\\
page title & 5.6 & 26 & - & -\\
section title & 3.3 & 22 & - & -\\
header & 19.7 & 729 & 0.032\% & 2\%\\
table & 549.1 & 20545 & 24.2\% & 65.3\%\\
all & 585.5 & 20605 & 27.3\% & 72.4\%\\ \hline
\end{tabular}
\end{table}

\subsection{Experimental Setup}

The performance of table retrieval methods is evaluated with  Mean Average Precision (MAP), Mean Reciprocal Rank (MRR) and Normalized Discounted Cumulative Gain (NDCG) at cut-off points 5, 10, 15, and 20.
To test significance, we use a two-tailed paired t-test and use \textdagger /\ddag\ to denote  significance levels at $p=$0.05, 0.005 respectively.

Based on Section \ref{select}, we have three strategies to calculate salience scores of items and three ways to construct items (as a list of columns, rows, or cells) from a table. We list all the methods settings in Table \ref{all_method}.  To obtain the salience scores,  we use fastText word embeddings~\cite{bojanowski2017enriching}.\footnote{\url{https://github.com/facebookresearch/fastText/}} Note that a different tokenization approach is used because fastText is not pre-trained on WordPiece tokenized corpus. We replace all non-numerical and non-alphabet characters with space and simply split sequences by space. 
Following the same experimental setup in \cite{zhang2018ad}, five-fold cross-validation is used when evaluating different methods.
We release our code on GitHub.\footnote{\url{https://github.com/Zhiyu-Chen/SIGIR2020-BERT-Table-Search}}

\begin{table}
\centering
\caption{The settings of all proposed methods, which use different item types and content selectors.}
\label{all_method}
\begin{tabular}{@{}lcc@{}}
\toprule
\multicolumn{1}{c}{Method Name} & Item type & Content Selector \\ \midrule
Hybrid-BERT-Row-Sum & Row & Sum Salience \\
Hybrid-BERT-Row-Mean & Row & Mean Salience \\
Hybrid-BERT-Row-Max & Row & Max Salience \\
Hybrid-BERT-Col-Sum & Column & Sum Salience \\
Hybrid-BERT-Col-Mean & Column & Mean Salience \\
Hybrid-BERT-Col-Max & Column & Max Salience \\
Hybrid-BERT-Cell-Sum & Cell & Sum Salience \\
Hybrid-BERT-Cell-Mean & Cell & Mean Salience \\
Hybrid-BERT-Cell-Max & Cell & Max Salience \\ \bottomrule
\end{tabular}
\end{table}

\begin{table*}
\centering
\caption{The superscript \textdagger\ shows statistically significant improvements for the method compared with all other methods.}
\label{result1}
\begin{tabular}{@{}lllllll@{}}
\toprule
\multicolumn{1}{c}{\begin{tabular}[c]{@{}c@{}}Method Name\end{tabular}} & \multicolumn{1}{c}{MAP} & \multicolumn{1}{c}{MRR} & \multicolumn{1}{c}{NDCG@5} & \multicolumn{1}{c}{NDCG@10} & \multicolumn{1}{c}{NDCG@15} & \multicolumn{1}{c}{NDCG@20} \\ \midrule
STR & 0.5711 & 0.6062 & 0.5762 & 0.6048 & 0.6102 & 0.6111 \\
Hybrid-BERT-text & 0.6003 & 0.6321 & 0.6023 & 0.6284 & 0.6322 & 0.6336 \\
Hybrid-BERT-Rand-Row & 0.6056 & 0.6356 & 0.6110 & 0.6294 & 0.6340 & 0.6350 \\
Hybrid-BERT-Rand-Col & 0.6105 & 0.6441 & 0.6094 & 0.6321 & 0.6388 & 0.6392 \\
Hybrid-BERT-Rand-Cell & 0.6124 & 0.6411 & 0.6117 & 0.6317 & 0.6381 & 0.6386 \\ \midrule
Hybrid-BERT-Cell-Mean & 0.6104 & 0.6364 & 0.6148 & 0.6337 & 0.6385 & 0.6388 \\
Hybrid-BERT-Cell-Max & 0.6129 & 0.6410 & 0.6166 & 0.6349 & 0.6391 & 0.6395 \\
Hybrid-BERT-Cell-Sum & 0.6207 & 0.6473 & 0.6227 & 0.6397 & 0.6450 & 0.6454 \\
Hybrid-BERT-Row-Mean & 0.6196 & 0.6490 & 0.6216 & 0.6406 & 0.6456 & 0.6463 \\
Hybrid-BERT-Row-Max & \textbf{0.6311} & \textbf{0.6673}\textsuperscript{\textdagger} & \textbf{0.6361} & \textbf{0.6519} & \textbf{0.6558} & \textbf{0.6564} \\
Hybrid-BERT-Row-Sum & 0.6199 & 0.6487 & 0.6168 & 0.6385 & 0.6436 & 0.6445 \\
Hybrid-BERT-Col-Mean & 0.6108 & 0.6395 & 0.6168 & 0.6340 & 0.6406 & 0.6412 \\
Hybrid-BERT-Col-Max & 0.6086 & 0.6324 & 0.6133 & 0.6297 & 0.6357 & 0.6362 \\
Hybrid-BERT-Col-Sum & 0.6131 & 0.6399 & 0.6131 & 0.6308 & 0.6384 & 0.6390 \\ \bottomrule
\end{tabular}
\end{table*}

\subsection{Baselines}

We implement the following baseline methods:
\begin{itemize}
    \item \textbf{Semantic Table Retrieval (STR)} This is the method proposed by Zhang and Balag~\cite{zhang2018ad} which is the previous state-of-the-art method. It first represents queries and tables in multiple semantic spaces. Then multiple semantic matching scores are calculated based on the representations of queries and tables. Pointwise regression using Random Forest is used to fit those semantic features combined with other features. Like the original STR implementation, we set the number of trees to 1000 and the maximum number of features in each tree to 3.
    \item \textbf{Hybrid-BERT-text} Only context fields are used and the table is not encoded except the table headers which are also considered as a context field. 
    \item \textbf{Hybrid-BERT-Rand-Col} Randomly selecting column items when constructing the BERT input. 
    \item \textbf{Hybrid-BERT-Rand-Row} Randomly selecting row items when constructing the BERT input.
    \item \textbf{Hybrid-BERT-Rand-Cell} Randomly selecting cells from the table when constructing the BERT input.
\end{itemize}
For the BERT-based methods, we use the features proposed in \cite{zhang2018ad} as $v_{a}$.

\subsection{Experimental Results}\label{rs1}

We  summarize our experimental results in Table~\ref{result1}. We can see that all BERT-based models can achieve better results than semantic table retrieval (STR).
Even without encoding the tables, Hybrid-BERT-text can still outperform STR, which demonstrates that BERT can extract informative features from tables and context fields for ad hoc table retrieval.  Randomly selecting columns, rows and cells have a marginal improvement on Hybrid-BERT-text, indicating that encoding the table content has the potential to further boost  performance.
In addition, the differences in performance among randomly selecting columns, rows and cells are not statistically significant.
The answer to \textbf{RQ1} is very straightforward: all BERT based models with different content selection methods can perform better than the previous state-of-the-art method. Though the gain of performance is statistically significant at $p=0.005$ level, BERT makes the main contribution, since only encoding context fields can achieve impressive results.

Next, we discuss the impact of item type and content selector.
Comparing the results in Table \ref{result1}, we observe that in general row item based methods are better than cell item based methods, and cell item based methods are better than column item based methods.
Among all the methods, Hybrid-BERT-Row-Max achieves the best results across all metrics compared with all other methods. The improvement over all other methods is statistically significant at 0.05 level for MRR, and statistically significant at 0.05 level for NDCG@5, NDCG@15 and NDCG@20 except for Hybrid-BERT-Cell-Sum.  
It means that selecting  rows that have the most significant signals is an effective strategy to construct BERT input within the length limit. 
In contrast to row items, column selection and cell selection based methods seem to be less effective. For several cases, content selection strategies for column/cell items even have worse performance than randomly selecting columns/cells. For example, Hybrid-BERT-Col-Max has MRR of 0.6324 while Hybrid-BERT-Rand-Col has MRR of 0.6441. Different from row items, max salience selector does not show superiority over other selectors for column items and cell items. 
It is expected that Hybrid-BERT-Rand-Col has better performance than Hybrid-BERT-Rand-Row, because a table is less likely to have more columns than rows, which means the probability of a potential optimal column to be selected is higher than that of a potential optimal row to be selected. 
For cell items, the sum salience selector shows marginally better performance than the other two selectors. And for column items, there is no clear best content selector but max salience selector seems to be the least effective.

\begin{table*}[h]
\centering
\caption{The setting of our methods where only BERT features are used.}
\label{result2}
\begin{tabular}{@{}lllllll@{}}
\toprule
\multicolumn{1}{c}{\begin{tabular}[c]{@{}c@{}}Method Name\end{tabular}} & \multicolumn{1}{c}{MAP} & \multicolumn{1}{c}{MRR} & \multicolumn{1}{c}{NDCG@5} & \multicolumn{1}{c}{NDCG@10} & \multicolumn{1}{c}{NDCG@15} & \multicolumn{1}{c}{NDCG@20} \\ \midrule
BERT-text & 0.5958 & 0.6240 & 0.5972 & 0.6206 & 0.6283 & 0.6287 \\
BERT-Rand-Row & 0.6005 & 0.6271 & 0.6063 & 0.6266 & 0.6310 & 0.6314 \\
BERT-Rand-Col & 0.6067 & 0.6400 & 0.6093 & 0.6327 & 0.6374 & 0.6380 \\
BERT-Rand-Cell & 0.6075 & 0.6358 & 0.6116 & 0.6287 & 0.6362 & 0.6369 \\ \midrule
BERT-Cell-Mean & 0.6056 & 0.6331 & 0.6017 & 0.6274 & 0.6340 & 0.6343 \\
BERT-Cell-Max & 0.5967 & 0.6275 & 0.6013 & 0.6209 & 0.6299 & 0.6307 \\
BERT-Cell-Sum & 0.6149 & 0.6436 & 0.6151 & 0.6345 & 0.6420 & 0.6424 \\
BERT-Row-Mean & 0.6055 & 0.6365 & 0.6064 & 0.6314 & 0.6358 & 0.6363 \\
BERT-Row-Max & \textbf{0.6277} & \textbf{0.6600} & \textbf{0.6274} & \textbf{0.6465} & \textbf{0.6517} & \textbf{0.6532} \\
BERT-Row-Sum & 0.6113 & 0.6302 & 0.6077 & 0.6307 & 0.6356 & 0.6370 \\
BERT-Col-Mean & 0.6026 & 0.6318 & 0.6079 & 0.6269 & 0.6334 & 0.6339 \\
BERT-Col-Max & 0.6095 & 0.6398 & 0.6109 & 0.6319 & 0.6379 & 0.6385 \\
BERT-Col-Sum & 0.6059 & 0.6257 & 0.6050 & 0.6260 & 0.6339 & 0.6343 \\ \bottomrule
\end{tabular}
\end{table*}

\begin{table*}[ht!]
\centering
\caption{Results using feature-based approaches. The superscript \ddag\  denotes statistically significant improvements over all baseline methods.}
\label{feature_based}
\begin{tabular}{@{}lllllll@{}}
\toprule
\multicolumn{1}{c}{Method Name} & \multicolumn{1}{c}{MAP} & \multicolumn{1}{c}{MRR} & \multicolumn{1}{c}{NDCG@5} & \multicolumn{1}{c}{NDCG@10} & \multicolumn{1}{c}{NDCG@15} & \multicolumn{1}{c}{NDCG@20} \\ \midrule
Hybrid-BERT-text & 0.6287 & 0.6546 & 0.6171 & 0.6489 & 0.6531 & 0.6536 \\
Hybrid-BERT-Rand-Col & 0.6590 & 0.6722 & 0.6481 & 0.6629 & 0.6692 & 0.6694 \\
Hybrid-BERT-Rand-Row & 0.6139 & 0.6418 & 0.6107 & 0.6345 & 0.6409 & 0.6411 \\
Hybrid-BERT-Rand-Cell & 0.6195 & 0.6554 & 0.6195 & 0.6382 & 0.6465 & 0.6466 \\ \midrule
Hybrid-BERT-Row-Max & \textbf{0.6737}\textsuperscript{\ddag}  & \textbf{0.7139}\textsuperscript{\ddag}  & \textbf{0.6633}\textsuperscript{\ddag}  & \textbf{0.6875}\textsuperscript{\ddag}  & \textbf{0.6924}\textsuperscript{\ddag}  & \textbf{0.6926}\textsuperscript{\ddag}  \\
Hybrid-BERT-Col-Mean & 0.6379 & 0.6582 & 0.6229 & 0.6449 & 0.6540 & 0.6542 \\
Hybrid-BERT-Cell-Sum & 0.6643 & 0.6806 & 0.6529 & 0.6686 & 0.6739 & 0.6740 \\ \bottomrule
\end{tabular}
\end{table*}

Three types of items are coherent units of the table with different granularities.
A cell is the smallest unit compared with a row item or a column item.  Usually, a column item is longer than a row item depending on the layout of the table. After manually examining some returned items, we find that cell item based methods are more biased towards returning items including query terms, while the methods based on the other two item types are forced to include some context information. Taking Figure \ref{cell_intent} as an example, all the returned items include the term ``wrestler'' which appears in the rightmost column that includes a list of short biographies of professional wrestlers. However, for row items, other context information such as the names of the wrestlers are forced to be included. Since column items are usually longer than row items, if the content selector fails to return the most relevant column item as the first one, 
the model is less likely to achieve good performance.
Based on our experiment results, we observe that max salience selector with row items has the best balance between accuracy and robustness, which answers \textbf{RQ3}.

\section{Discussion}

In this section, we continue the discussion of our proposed methods. 

\subsection{Ranking Only with BERT}\label{only_bert}

To answer \textbf{RQ2}, we run the experiments that only use BERT features which means $f$ equals $f_{bert}$ in Equation \ref{fusion}. The results are shown in Table~\ref{result2} where the method names correspond to the ones in Table~\ref{result1} except the STR baseline and the prefix ``Hybrid-'' is removed. In all cases, performance decreases slightly when additional features are not used.
In answer to \textbf{RQ2},  without additional features, all the proposed methods including baselines can outperform STR. Even without encoding table content, BERT-text can still achieve good performance which means the context fields are very important for ad hoc table retrieval.
The conclusions are consistent with Section \ref{rs1}: sum salience selector is the best for cell items and max salience selector with row items still performs the best when only BERT features are used.

\subsection{Generalization to Another Domain}
Though we conclude that the max salience selector with row items is the best method, the conclusion may depend on the corpus. Therefore, we also conduct experiments on the dataset from another domain. To do this, we use an open-domain dataset WebQueryTable\footnote{\url{https://github.com/tangduyu/Table-Intelligence/tree/master/table-search}} introduced by Sun et al.~\cite{sun2019content}. Unlike WikiTables where all the tables are from Wikipedia, the tables in WebQueryTable are collected from queried web pages returned by a commercial search engine. In total,  21,113 query-table pairs are manually annotated and the dataset is pre-split into training (70\%), development (10\%) and test (20\%). In this scenario, no additional features are available for this corpus so  only  BERT features are used. Additionally, table caption, sub-caption and headers are used as context fields. The preprocessing is the same with WikiTables. We do not use the development set since we do not search for hyperparameters.
We calculate the MAP scores of our models which are also reported by Sun et al.~\cite{sun2019content}. The results of the best BERT baseline method and the proposed method are shown in Table \ref{WQT}.
The final results are also consistent with conclusions in Section \ref{rs1}---that max salience selector with row items is the best strategy.\footnote{We did not reproduce their method. We assume the results are comparable since the dataset is pre-split.} 
Therefore, we can see that training BERT on row items with max salience selector is also an effective strategy for datasets in other domains, which makes the answers to \textbf{RQ2} and \textbf{RQ3} more convincing.

\subsection{Feature-Based Approach of BERT}

In Section \ref{rs1}, we use the fine-tuning approach that jointly fine-tunes the whole framework. In the experiment, we tried different methods to incorporate additional features. For example, we can directly concatenate additional features without any transformation with BERT features and feed the concatenated vector to the regression layer. We also tried to predict two relevance scores with BERT features and additional features separately, and then linearly transform them into  a weighted relevance score. However, all of those variants perform worse than BERT-text.  It is possible that BERT performance highly depends on the optimization strategy and adding other components for joint training can have negative impact on the fine-tuning process. 
To avoid such a case, we adapt BERT to a feature-based approach. First we use the fine-tuning approach to train BERT without additional features like in Section \ref{only_bert}. Then we optimize the whole framework as in Section \ref{rs1} except that BERT weights are not updated. The results are shown in Table \ref{feature_based}. For the three item types, we only include the results of models using the best content selectors. 
All methods have significant improvements compared with fine-tuned approaches.  Among the baselines, Hybrid-BERT-Rand-Col has the most improvement, which is even better than the best performance of BERT using content selectors for column items. Hybrid-BERT-Row-Max still achieves the best performance and the improvements over baselines are statistically significant at the level of $p=0.005$. 

So far, we observe that max salience selector with row items is the best strategy to construct inputs for BERT. In the feature-based approaches, it is more obvious that sum salience selector is the best one for cell items and mean salience selector is the best one for column items. 

\begin{table}
\centering
\caption{Results on WebQueryTable dataset.}
\label{WQT}
\begin{tabular}{@{}ll@{}}
\toprule
\multicolumn{1}{c}{Method Name} & \multicolumn{1}{c}{{MAP}} \\ \midrule
{Feature + NeuralNet \cite{sun2019content}} & {0.6718} \\
{BERT-Rand-Cell} & {0.6414} \\
BERT-Row-Max & \textbf{0.7104} \\ \bottomrule
\end{tabular}
\end{table}

\section{Analysis of BERT Features}
\begin{figure*}[h]
\centering
\includegraphics[width=\textwidth]{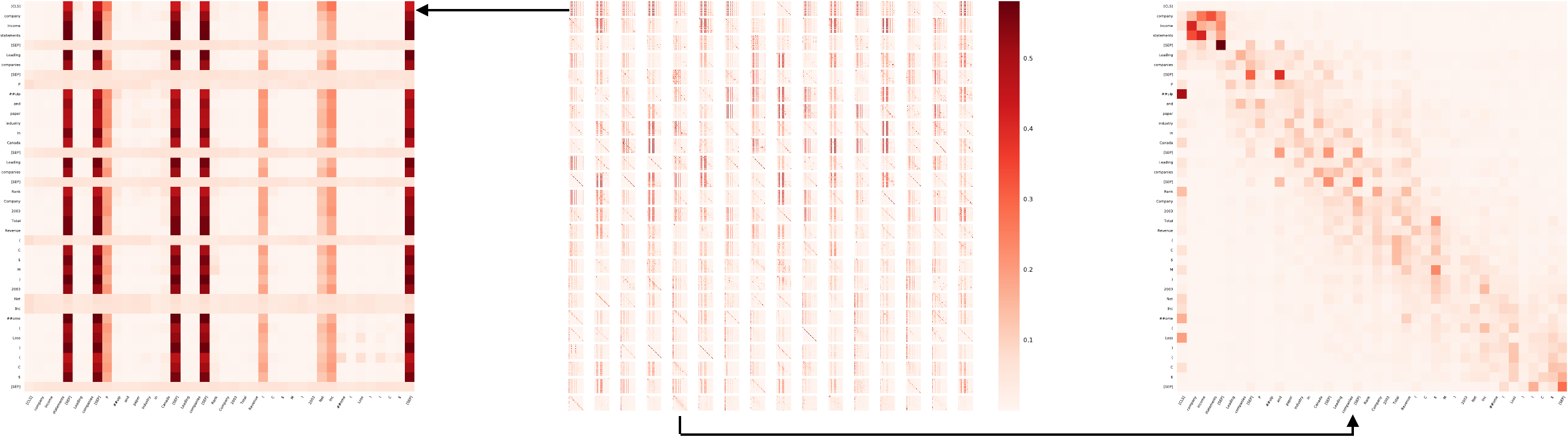}
\caption{Middle figure includes all attention maps of a random test set example.
Left figure shows the attention map of head 1 in the last layer. Similar attentions for [SEP] tokens result in the grid-look. Right figure shows the attention map of head 5 in the 1st layer with intra-sequence attention pattern. Attention weights with larger absolute values have darker colors. }
\label{attn}
\end{figure*}

Though BERT achieved new state-of-the-art results on various tasks, it is still unclear what are the exact mechanisms behind its success. In this section, we dive into the analysis of BERT for the table retrieval task. 
For illustration purposes, the results presented in this section are based on the weights of BERT-Row-Max. However, we observe similar patterns among different BERT-based methods and therefore the conclusions can also be applied to other methods.

\subsection{Self-Attention Patterns}

Compared to general scenarios where BERT is used for single-sequence or sequence-pair tasks, there are more than two sequences involved in the input of BERT for the table retrieval task and the sequence could have a lot of [SEP] tokens. BERT practitioners know [SEP] is a special token that is used as a delimiter of sequences.  For our case, there could be a lot of [SEP] tokens in a single input and the number of [SEP] tokens are different across different samples. 
In this section we explore whether the self-attention patterns of BERT used in this paper which involve multiple sequences are different from a BERT model fine-tuned on single sequence/sequence pair tasks.  

We draw all the attention maps of a random example from the test set in Figure \ref{attn}.
We find all the types of self-attention maps categorized in \cite{kovaleva-etal-2019-revealing}: vertical, diagonal, vertical with diagonal, block and heterogeneous.
We find that  [SEP] embeddings in lower layers are attended or attending more differently than those in higher layers. Taking the 1st self-attention head in the 4th layer as an example, the 1st [SEP] embedding mainly attends to itself, while the other [SEP] embeddings mainly attend to [CLS] embedding.  In contrast, the attentions for [SEP] tokens are very similar in higher layers resulting in a lot of grid-like attention maps (head 1 in the last layer as shown in Figure \ref{attn}). 
We also quantitatively measure the embeddings of different [SEP] tokens and calculate the smallest cosine similarity among all pairs of [SEP] in the same layer.  The smallest cosine similarity is 0.78 in the 1st layer but increases close to 1 in higher layers, which means [SEP] tokens have different embeddings in lower layers, and after layers of self-attention, they have almost the same representations.

Besides the types of attention maps described by Kovaleva et al.~\cite{kovaleva-etal-2019-revealing}, we observe some attention maps that look like scatter plots, which include sparse small blocks (e.g., head 1 in the 4th layer).
This is because multiple sequences are included in a single input separated by [SEP] and some attention heads have a strong preference to put attention on multiple sequences (inter-sequence attention). 
We also observe there are self-attention heads
that show intra-sequence attention patterns. For example, caption and section title both attend to themselves a lot in head 9 of the 1st layer. Query tokens attend a lot to themselves in head 5 of the 1st layer (right in Figure \ref{attn}). 
The existence of \textbf{intra-sequence} and \textbf{inter-sequence attention patterns} may indicate that BERT can learn various sequence-level features through self-attention.

\begin{figure}[h]
\centering
\includegraphics[width=0.475\textwidth]{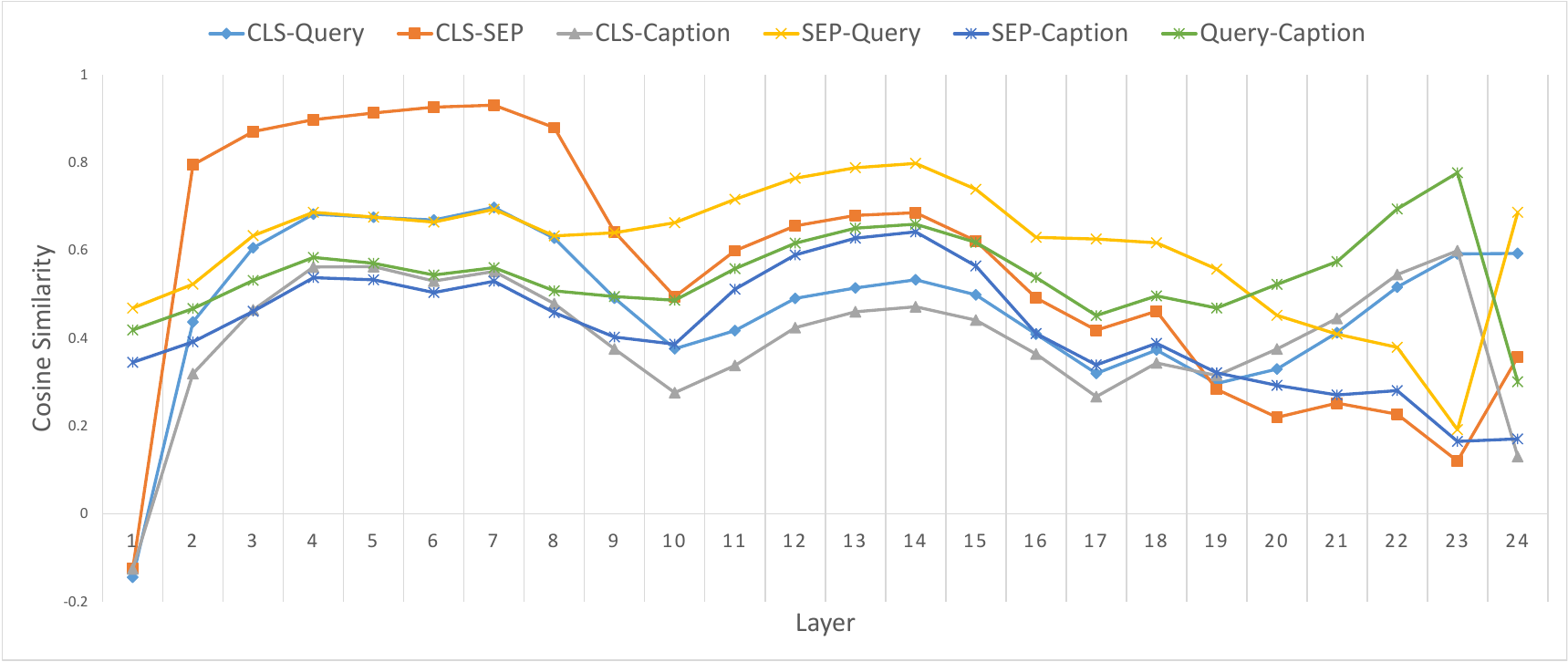}
\caption{Average cosine similarity among different types of tokens in different layers. }
\label{tokens}
\end{figure}

\subsection{BERT Embedding Comparison}
In the experiments, only the [CLS] embedding in the last layer is used as BERT features and the rest are not utilized. Here we further analyze the relationships among different types of BERT embeddings. 

For each sample in the testing set, we extract embeddings corresponding to query tokens and average them as the query representation. We do the same for [SEP] and caption. Then we calculate the cosine similarity between every two of [CLS], [SEP], query and caption. 
We show the average cosine similarity of testing samples at different layers in Figure \ref{tokens}. We observe that the patterns between special/query tokens and table/context field tokens are similar, which means in Figure \ref{tokens}, if we replace caption with other context fields or selected items, the general patterns do not change. For example, ``Query-Caption'' is similar to ``Query-Page title'' and ``CLS-Caption'' is similar to ``CLS-Page title''.

In the 1st layer, [SEP] is close to query and caption while [CLS] is far from [SEP], query and caption. From layer 2 to layer 8, we note that [CLS] is very close to [SEP], which may indicate that [CLS] aggregates segment-level information through these layers. In contrast,  the similarities among [SEP], query and caption do not change significantly from layer 2 to layer 14. 
It is interesting that from layer 23 to the last layer, query and [CLS] become closer but far away from caption. In the last layer, [SEP] is closer to query than [CLS], which may indicate [SEP] captures more query features than [CLS]. 


\section{Conclusion}

We have addressed the problem of ad hoc table retrieval with the deep contextualized language model BERT. Considering the structure of a table, we propose three content selectors to rank table items in order to construct input for BERT which effectively utilize useful information from tables and overcome the input length limit of BERT to some extent.  
We combine BERT features and other tables features to solve the table retrieval task as a pointwise regression problem. Our proposed Hybrid-BERT-Row-Max method outperforms the previous state-of-the-art and BERT baselines with a large margin on WikiTables dataset. Through empirical experiments, we find that using the max salience selector with row items is the best strategy to construct BERT input. Overall, we also find that sum salience selector is the best for cell items. While for column items, mean salience selector only seems to be the best when a feature-based approach is used.
We further show that the feature-based approach of BERT is better than jointly training BERT with a feature fusion component. 
We also conduct experiments on WebQueryTable dataset and demonstrate that our method generalizes to other domains. 

Our analysis on fine-tuned BERT shows that various sequence-level features are captured by the self-attention of BERT and [CLS] embedding tends to aggregate sequence-level information, which could explain why using it as features is effective for the ad hoc table retrieval task.  
We also find that [SEP] embeddings from the last layer of BERT are very close to query embeddings, which indicates that making use of [SEP] has the potential to further improve the performance. 
Though the motivation behind this paper is that different content selection strategies should be used for different queries, we do not explore how to design a model to choose the best selector. In fact, it is possible that for different types of queries, we should choose different content selector.
Future work could design a framework that automatically chooses the strategy considering the query types. 
Besides, designing pretraining tasks for tables and pretraining BERT on a large table collection could be promising to further improve the performance of BERT on table-related tasks such as table retrieval.

\begin{acks}
This material is based upon work supported by the National Science Foundation under Grant No.\ IIS-1816325. The authors would like to thank the anonymous reviewers for valuable comments, and Ao Luo and Shengfeng Pan from Shenzhen Zhuiyi Technology Co., Ltd. for useful discussion about BERT.
\end{acks}
\newpage
\bibliographystyle{ACM-Reference-Format}
\bibliography{acmart.bib}

\appendix

\begin{figure*}[t]
\centering
\includegraphics[width=0.9\textwidth]{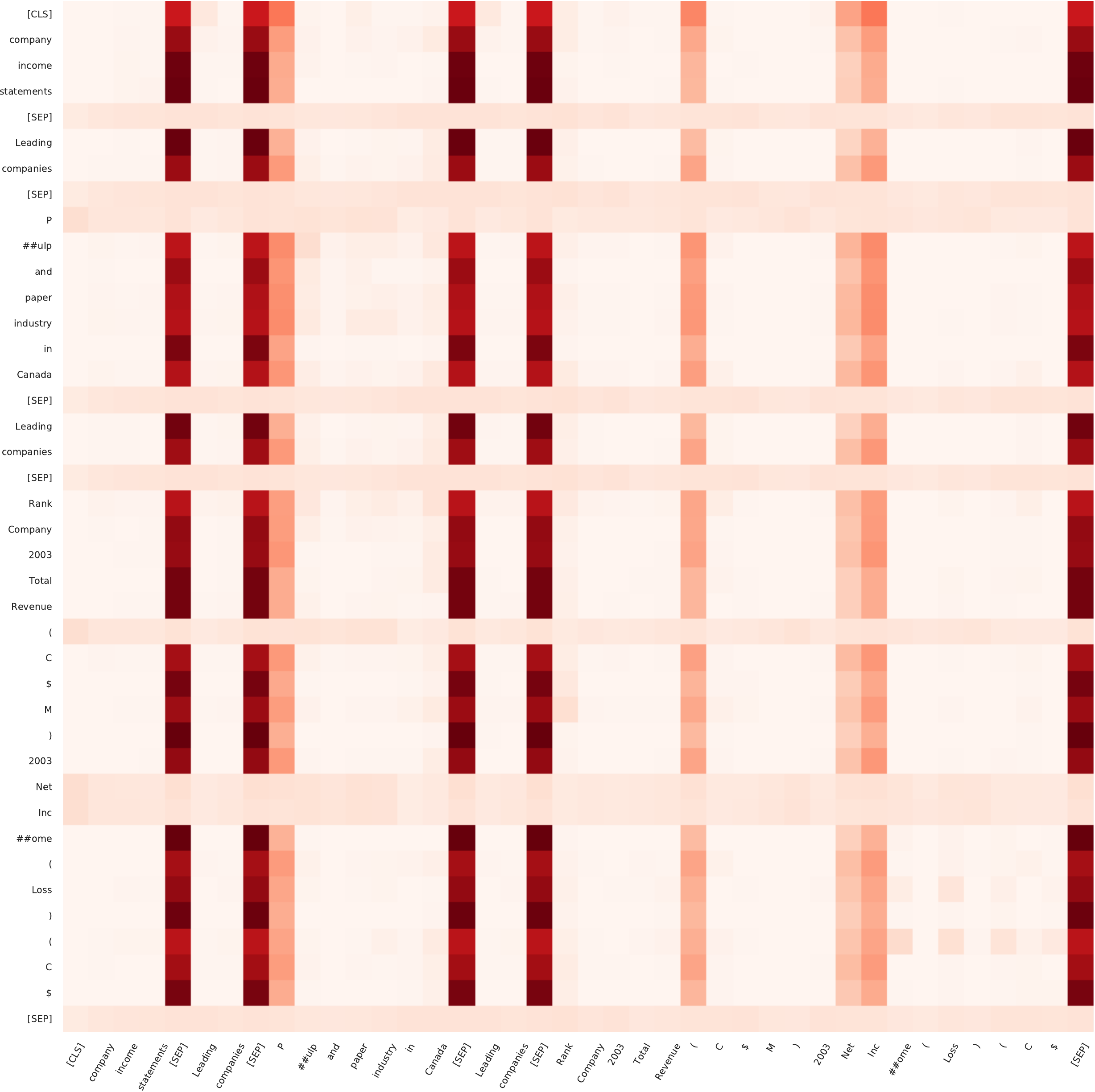}
\caption{The attention map of 1st attention head at last layer. }
\end{figure*}

\section{Attention Maps}

Here we list all the attention maps mentioned in Figure \ref{attn}.

\begin{figure*}[h!]
\centering
\includegraphics[width=0.9\textwidth]{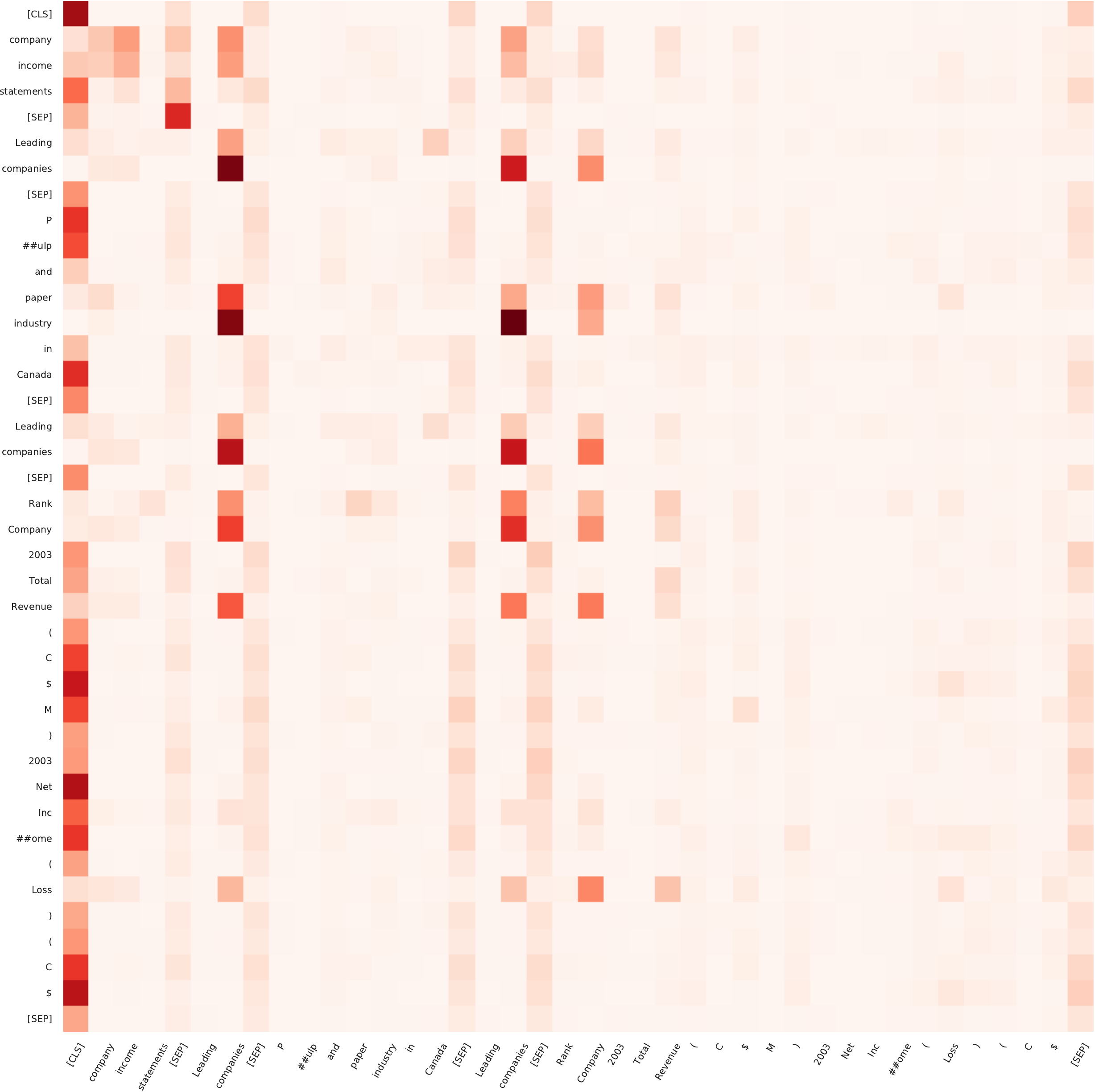}
\caption{The attention map of 1st attention head at 4-th layer. }
\end{figure*}

\begin{figure*}[h!]
\centering
\includegraphics[width=0.9\textwidth]{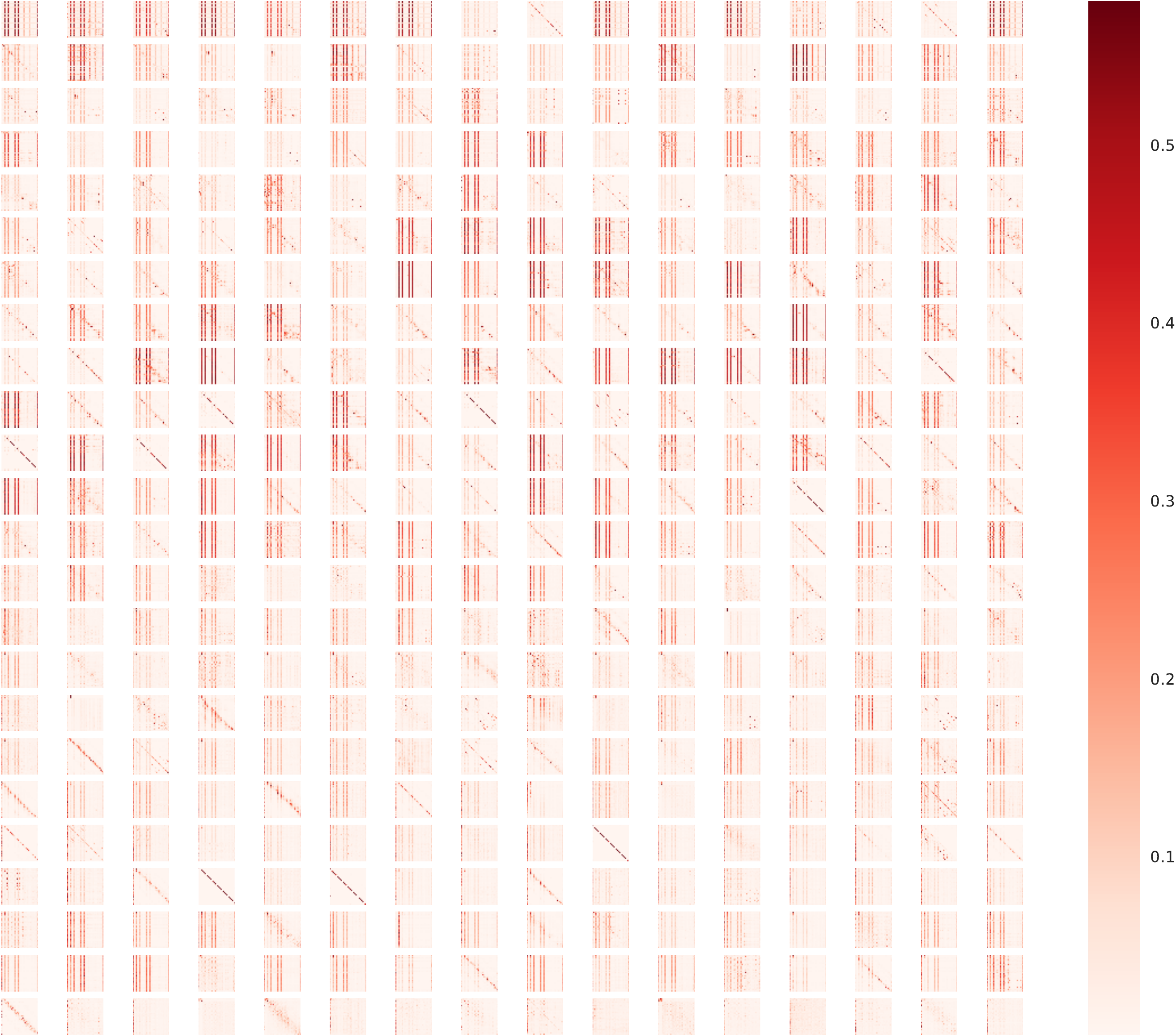}
\caption{The attention maps of all attention heads. }
\end{figure*}

\end{document}